\documentclass[%
reprint,
prl,
twocolumn,
groupedaddress,
amsmath, amssymb, aps,raggedbottom
]{revtex4}
\usepackage{amsmath}
\usepackage{amssymb}
\usepackage{graphicx}
\usepackage{epsfig}
\usepackage{epstopdf}
\usepackage{dcolumn}
\usepackage{bm}
\usepackage{natbib}

\newcommand{\ket}[1]{\ensuremath{\left|#1\right\rangle}}
\begin{document}

\title{Coherent Quantum Transport in Photonic Lattices}

\author{Armando Perez-Leija$^{1,\dag}$} \author{Robert Keil$^{2,\dag}$} \author{Alastair Kay$^{3,4,\dag}$} \author{Hector Moya-Cessa$^{1,5}$} \author{Stefan Nolte$^{2}$} \author{Leong-Chuan Kwek$^{4,6}$} \author{Blas M. Rodr\'iguez-Lara$^{5}$} \author{Alexander Szameit$^{2}$} \author{Demetrios N. Christodoulides$^{1,}$}\email{demetri@creol.ucf.edu}
\affiliation{$^{1}$CREOL$/$College of Optics, University of Central Florida, Orlando, Florida, 32816, USA\\$^{2}$Institute of Applied Physics, Friedrich-Schiller Universit\"{a}t Jena, Max-Wien-Platz 1, 07743 Jena, Germany\\$^{3}$Keble College, Parks Road, University of Oxford, Oxford, OX1 3PG, UK\\$^{4}$Centre for Quantum Technologies, National University of Singapore, Singapore 117543\\$^{5}$INAOE, Coordinaci\'{o}n de \'{O}ptica, Luis Enrique Erro No.1, 72840 Toantzintla, Pue., M\'{e}xico\\$^{6}$Institute of Advanced Studies (IAS) and National Institute of Education, Nanyang Technological University, Singapore 639673\\$^{\dagger}$These authors contributed equally.
}
\begin{abstract}
Transferring quantum states efficiently between distant nodes of an information processing circuit is of paramount importance for scalable quantum computing. We report on the first observation of a perfect state transfer protocol on a lattice, thereby demonstrating the general concept of transporting arbitrary quantum information with high fidelity. Coherent transfer over 19 sites is realized by utilizing judiciously designed optical structures consisting of evanescently coupled waveguide elements. We provide unequivocal evidence that such an approach is applicable in the quantum regime, for both bosons and fermions, as well as in the classical limit. Our results illustrate the potential of the perfect state transfer protocol as a promising route towards integrated quantum computing on a chip. 
\end{abstract}

\pacs{42.50.Ex, 05.60.Gg, 42.82.Et}

\maketitle
Quantum computers promise unprecedented levels of computational power over those anticipated from classical systems \cite{Deutsch,Grover,Shor}. To fulfill this potential, a key milestone in the development of quantum computing is the coherent transfer of states between numerous qubits in an extended circuit. A major challenge therein is that typically the actual carriers of information do not physically move, irrespective of whether the computational devices are implemented in ionic \cite{Cirac,Kielpinski,Kim}, solid state \cite{Berezovsky,Hanson} or superconducting systems \cite{Yamamoto,Neeley}. Although there are suggestions for moving ions \cite{Home}, this concept usually leads to substantial complications and may not be feasible in many settings. Hence, the transfer of quantum states across a static information system is nowadays considered by many as the protocol of choice on these platforms.

The efficacy of any transfer procedure is measured by the fidelity F, with perfect transfer corresponding to F=1. In a classical (\emph{incoherent}) protocol, the best transfer possible can be achieved by first measuring the state, and subsequently communicating the result, thus allowing reconstruction of the initial state at a distant site. In this case, the fidelity can never exceed the well-known limit of 2/3 or 67\%. In order to surpass this bound, the transport protocol must demand that coherence should be maintained throughout the transfer process. A straightforward approach to satisfy this latter requirement is to use a sequence of gates capable of switching adjacent qubits (so called SWAP gates) \cite{Nielsen}. In theory, the short-range interaction in such architectures is sufficient to support long-range coherent transport. In reality however, apart from practical issues pertaining to the control of a large number of distinct SWAP gates, the effects from even minute imperfections tend to accumulate, thus resulting in a drastic degradation of the quality of the input state. To illustrate the extent of the aforementioned challenge, even if the efficiency of a single gate is 98\%, after a sequence of only twenty such gates, the quality of the input state will be degraded below the classical threshold.

Recent theoretical advances have demonstrated that if coherence can be maintained across many qubits, the transfer of quantum states can be obtained much faster, more robustly, and with less active intervention \cite{Nielsen}. Indeed, such a protocol can achieve high-fidelity transfer by merely manipulating the coupling mechanism between adjacent qubits in a chain. In such an architecture, it is sufficient to pre-engineer the interaction Hamiltonian so that the intrinsic dynamics themselves facilitate the transfer of the state. The only action one needs to impose on the system can be performed ahead of the transfer process, thus enabling the minimization of detrimental couplings to the environment. In other words, after supplying the state at the input port, it just has to be retrieved from the output.

Initial proposals concentrated on evaluating the efficacy of a chain of spins subject to a uniformly coupled Heisenberg Hamiltonian \cite{Bose}. For such  Hamiltonians, perfect quantum state transfer is only possible for 2 or 3 qubits \cite{Christandl1}. Subsequently it was found that perfect state transfer can be achieved even for arbitrarily long chains provided the couplings between neighboring sites can be appropriately engineered \cite{Christandl2}. Thereafter, a plethora of theoretical results have described how these transfer protocols could be implemented in every conceivable scenario (see for example \cite{Kay,Bose2} and references therein). However to date, experimental realizations of such schemes have only been reported in the token case of a chain of 3 qubits using magnetic resonance \cite{Zhang}.

\begin{figure}[h!]
\begin{center}
\includegraphics[width=3.18in]{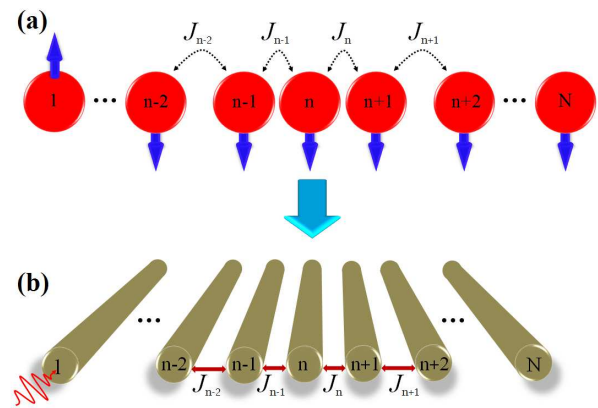}
\caption{(Color online) Parallel correspondence between (a) Heisenberg spin chains and (b) waveguide arrays. In (a) spin-$1/2$ particles in the state $\ket{\uparrow\downarrow\ldots\downarrow}$ involving nearest-neighbor interactions $J_{n}$. In (b) an array of optical waveguides with evanescent nearest-neighbor coupling $J_{n}$. In (a) the vertex spin $1$ has been flipped up whereas in the waveguide system (b) it is represented by photons being launched into the first waveguide element.}
\label{f1}
\end{center}
\end{figure}

Here, we report an experimental demonstration of a genuine long-range coherent transport. We generalize the perfect quantum state transfer to another physical platform: light in evanescently coupled waveguides, so-called photonic lattices \cite{Christodoulides}. In fact, different configurations of optical waveguides have been employed in several investigations for the realization of quantum circuits and simulations of quantum walks \cite{Politi,Crespi,Sansoni}. Our proposed mechanism has a one to one correspondence with that in a spin chain; each qubit is represented by a distinct lattice site, in our case the individual waveguides, and the presence or absence of a photon at a given site corresponds to the  $\ket{1}$ and $\ket{0}$ states of the qubit.  Importantly, the carriers of information, in our arrangement the individual waveguide elements, remain static during the transport process. A major advantage of our approach is that the time evolution of the qubits is mapped onto a spatial coordinate along the waveguides, allowing a direct observation of evolution dynamics. We measure a transfer fidelity of 84\% through a system of 19 waveguides, thus proving the existence of long-range coherence in this optical array network. Even though no information is encoded in the photons themselves or in their quantum statistics, the underlying dynamics in these fully photonic lattices are formally identical to those occurring in a spin state transfer configuration. In addition, we study two-photon correlations, exhibiting bunching and anti-bunching behavior, thereby highlighting the differences between a bosonic and a fermionic state transfer system arising in the quantum regime.\\
In general, perfect coherent quantum transport requires a lattice of   coupled qubits described by the fermionic spin Hamiltonian \cite{Bose}:
\begin{equation}
\label{eq1}
H=\frac{1}{2}\sum_{n=1}^{N-1}J_{n}\left(X_{n}X_{n+1}+Y_{n}Y_{n+1}\right),
\end{equation}
where $X_{n}$ and $Y_{n}$ represent the Pauli matrices acting on qubit $n$, $N$  is the total number of sites or qubits involved in the spin chain, and the hopping parameter $J_{n}$ denotes the rate at which an excitation could couple from one site to another (see Figure \ref{f1}(a)). 
In this spin system, the probability amplitude $\alpha(t)$ associated with qubit n evolves in time according to the Schr\"{o}dinger equations 
\begin{equation}
\label{eq2}
\begin{split}
i\frac{d\alpha_{1}}{dt}&=J_{1}\alpha_{2},\\
i\frac{d\alpha_{n}}{dt}&=J_{n}\alpha_{n+1}+J_{n-1}\alpha_{n-1},\\
i\frac{d\alpha_{N}}{dt}&=J_{N-1}\alpha_{N-1},
\end{split}
\end{equation}
$(\hbar=1)$. The condition for perfect state transfer after time $t_{f}$ implies that $|\alpha_{N+1-n}(t_{f})|=|\alpha_{n}(t_{0})|$ and can only be achieved provided that $J_{N-n}=J_{n}$ \cite{Kay}. In fact, equidistant spacing of the eigenvalues of the Hamiltonian by integer multiples of $\pi/t_{f}$ is a direct consequence of this latter requirement \cite{Kay}. Based on these fundamental principles, one can relate the spin Hamiltonian of equation (1) to the \emph{x}-component of the angular momentum rotation matrix of a spin $(N-1)/2$ particle, resulting in the coupling condition $J_{n}=\pi\sqrt{n(N-n)}/2t_{f}$ \cite{Christandl2}. Any initial one-site excitation state is perfectly transferred from qubit $n$ to $N-n+1$ after a time $t_{f}$, and experiences perfect revivals after $2t_{f}$, up to a global phase. This specific set of hopping parameters has been considered in numerous contexts \cite{Christandl2,Bose2,Gordon,leijaX}, with potential applications outlined in yet more \cite{Peres,Kay2}. Even more importantly, it also turned out that the Hamiltonian (1) along with the coupling condition is the quintessential example of a perfect state transfer since it optimizes a variety of parameters \cite{Kay,Yung}. For instance, the transfer in this chain is robust to imperfect timing, that is, the fidelity of the transport is only marginally degraded at some deviation from  $t_{f}$. This robustness makes this arrangement superior to SWAP gates, where the fidelity can drop to zero even at a small deviation from the transfer time. Additionally, for a given maximum coupling strength, a chain designed based on the $J_{n}$ coupling condition is known to exhibit the shortest possible transfer time, which is twice as fast as a sequence of SWAP gates of the same maximum strength \cite{Yung}. A perfect state transfer in such a time unequivocally proves the presence of long-range coherence for timescales on the order of  $t_{f}$. 

Although the Hamiltonian (1) was originally proposed for fermionic qubits, its structure suggests that it could also be physically realizable in bosonic chain arrangements. In this work, we have implemented such a system using photonic lattices, where the coherent transport of light exhibits identical intrinsic dynamics as in fermionic spin chains. The formal analogy between these two systems is illustrated in Figures 1 (a) and (b). To this end, we use the aforementioned array of evanescently coupled waveguides obeying the parabolic distribution for the coupling coefficients between nearest-neighbor elements. In these waveguides, each photon evolves independently along the waveguides \cite{Bromberg}, obeying a set of Heisenberg equations that are entirely analogous to equations (2) except that here the creation operators $a_{n}^{\dagger}$ (as opposed to probability amplitudes) now evolve along the spatial propagation coordinate $Z$ in every waveguide. Hence, in order to achieve perfect state transfer in this configuration, the corresponding coupling matrix must follow the angular momentum rotation matrix, i.e., $\left(J_{x}\right)_{m,n}=f(n)\delta_{n,m+1}+f(n-1)\delta_{n,m-1}$ with $f(n)=\pi\sqrt{n(N-n)}/2z_{f}$, whereby in our case $z_{f}$ represents the distance for perfect transfer.

The eigenvectors of this particular Heisenberg spin lattice can be analytically obtained and are given by
\begin{equation}
\label{eq3}
\begin{split}
u_{n}^{m}=\left(\frac{2z_{f}}{\pi}\right)^{-\frac{1}{2}\left(N+1\right)+n}&\sqrt{\frac{\left(n-1\right)!\left(N-n\right)!}{\left(m-1\right)!\left(N-m\right)!}}\\
&\times P_{n-1}^{\left(m-n,N-m-n+1\right)}\left(0\right),
\end{split}
\end{equation}
where the functions $P_{n-1}^{\left(m-n,N-m-n+1\right)}\left(0\right)$ represent Jacobi polynomials of order $(n-1)$, evaluated at the origin. The eigenvalues $\lambda_{m}$ are distributed equidistantly within the interval $\left[-\pi\left(N-1\right)/2z_{f}, \pi\left(N-1\right)/2z_{f}\right]$ in steps of $\pi/z_{f}$. Using the eigenvectors and the corresponding eigenvalues, one can then obtain the probability amplitudes, over the entire lattice at distance $Z$, for any single photon excitation, $\Psi\left(Z\right)=\sum_{r=1}^{N}C_{r}u^{(r)}\exp\left( i\lambda_{r}Z\right)$, where $C_{r}=\left(u^{(r)}\right)^{\dag}\cdot \Psi\left(0\right)$. In general, the input-output states are related through the evolution matrix, $a_{p}^{\dagger}\left(0\right)=\sum_{n=1}^{N}T_{p,n}^{*}\left(Z\right)a_{n}^{\dagger}\left(Z\right)$, with $T_{p,n}^{*}\left(Z\right)$ denoting the Hermitian conjugate of the $\left(p, n\right)$ matrix element within the unitary transformation 
\begin{equation}
\label{eq4}
T_{p,q}\left(Z\right)=\sum_{r=1}^{N}u_{q}^{(r)}u_{p}^{(r)}\exp\left(i\lambda_{r}Z\right).
\end{equation}
The probability of detecting a photon at waveguide $p$, when launched at $q$, is given by the photon density $P_{p,q}\left(Z\right)=\langle a_{p}^{\dagger}a_{p}\rangle=\mid T_{p,q}\left(Z\right)\mid^{2}$. Since at integer values of revival distances $Z=2z_{f}s$ ($s$ being an integer) the matrix elements collapse to $T_{p,q}=e^{i\phi}\delta_{p,q}$, then $P_{p,q}$ indicates that revivals of probability will periodically occur in these systems irrespective of the total number of waveguide elements contained in the array or the initial site of excitation. On the other hand, at $Z=2z_{f}$ the unitary transformation leads to $T_{p,q}\left(Z=2z_{f}\right)=\pm \delta_{p,q}$, with the upper sign $+1$ corresponding to $N$  being an odd number while the lower sign $-1$ to $N$ being even. In other words, if the eigenvalues are odd multiples of $\pi/2z_{f}$ ($N$ even)  any initial state will exhibit perfect revivals at distances that are multiples of $Z=4z_{f}$ , whereas for eigenvalues being even multiples of $\pi/2z_{f}$ ($N$ odd) the states will spatially revive at integer multiples of $Z=2z_{f}$. Therefore, any one-site excitation state $\ket{\psi_{in}}=\ket{0, \ldots, 1_{n},\ldots,0}$ will be perfectly transformed (or transferred) into the state $\ket{\psi_{out}}=\ket{0, \ldots, 1_{N-n+1},\ldots,0}$ after a distance $z_{f}$. For example, when a single photon is launched into waveguide $n=1$, Eq.\eqref{eq4} implies that the fidelity of detecting it at waveguide $n$ is given by
\begin{equation}
\label{eq44}
F_{1,n}={N-1 \choose n-1}\left[\cos\left(\frac{\pi Z}{2z_{f}}\right)\right]^{2(N-n)}\left[\sin\left(\frac{\pi Z}{2z_{f}}\right)\right]^{2n-2}.
\end{equation} 
Interestingly, the single-photon approach even works in the regime of many photons - each photon must independently be transferred through the lattice provided that long range coherence is present in the system. In this vein, perfect state transfer for optical excitations can therefore be achieved also in the case of purely classical light.\\
In order to perform our experiments, we have implemented such spin-inspired waveguide arrays in bulk fused silica by employing direct femtosecond-laser inscription \cite{Szameit}. The coupling coefficients $J_{n}$ depend directly on the inter-waveguide separation $d_{n}$. Hence, the required parabolic coupling distribution can be achieved by choosing $d_{n}$ accordingly. Using the parameters given in appendix A, we inscribed a photonic lattice with $N=19$ waveguide elements, having a length of $L = 10 cm$. Linearly polarized light at $\lambda=633nm$ was injected into the lattice and was indirectly observed in the sample using fluorescence microscopy (Fig. 2). 

\begin{figure}[b!]
\begin{center}
\includegraphics[width=3.18in]{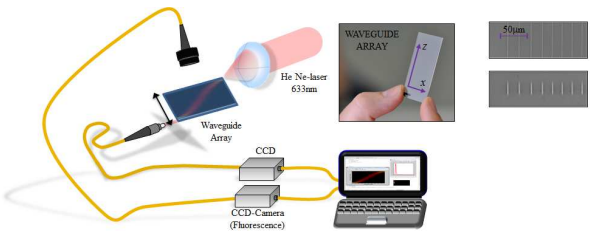}
\caption{(Color online) Experimental setup: Light from a 633nm laser source is coupled into the waveguide array. The intensity evolution is observed from the top via fluorescence from color centers, whereas the output intensity distribution is directly imaged onto a charged-coupled device (CCD).}
\label{f2}
\end{center}
\end{figure}

In Figure 3(a) we present the experimental demonstration of perfect state transfer over $19$ lattice sites, when light is launched into waveguide $n=1$ and coupled out of waveguide $n=19$. The simulations [Fig. 3(e)] fully confirm our observations. These results clearly demonstrate the coherent character of the long-range transport of photon-encoded qubits which are initialized into the relevant waveguide elements (acting as qubits). Quantitatively, the transfer fidelity \cite{fidelity} over the entire lattice is found to be 82\% at a transfer distance of $z_{f}=94mm$ , i.e., 82\% of the output light is observed in the intended waveguide. This value is below that anticipated from perfect transfer due to a variety of effects, but is nevertheless well in excess of the classical probability of success, 67\% (see appendix B for an error analysis). In this vein, transfers over arbitrarily long distances can be implemented, just by increasing the transverse size of the array. As all waveguides are identical, and merely the coupling varies across the lattice, an increase of the system size has no influence on the single-mode property of the individual guides. A striking feature of perfect state transfer offered by the spin Hamiltonian, is that, an input state not only can be transferred from qubit $1$ to $N$, but also from any other qubit $n$ to $N-n+1$ , i.e., perfect transport is not necessarily constrained to the two boundaries of the lattice. We experimentally demonstrate this process in Figs. 3 (b) and (d) where light is launched into waveguides $2$, $18$ and $19$ and is retrieved from the output at waveguides $18$, $2$ and $1$. For the non-boundary excitation the transfer fidelity was found to be 72\% and 74\%, respectively, whereas it reaches 84\% for the $1 \leftrightarrow 19$ excitation, surpassing the  classical threshold in each case. In all cases, our experimental data is fully supported by simulations, shown in Figs. 3 (f) to (h). Note that the primary physical reasons for the observed deviations from the ideal behavior are merely due to positioning and excitation inaccuracies, whereas full coherence is maintained (see appendix B). Furthermore, since propagation losses in our system are approximately 0.5 dB/cm and because they can be as low as 0.05 dB/cm \cite{Fukuda} such waveguide configurations are actually suitable for single photon experiments. 

\begin{figure}[t!]
\begin{center}
\includegraphics[width=3.18in,height=3.34in]{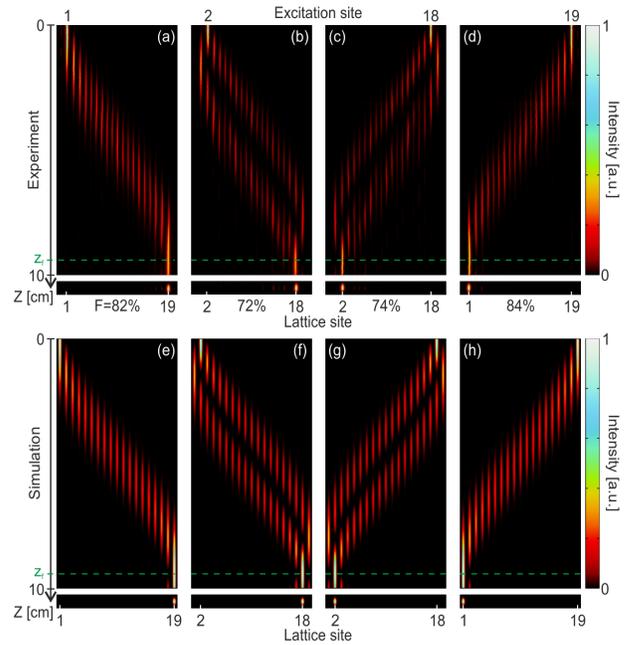}
\caption{(Color online) Transfer of a single-site excitation (a-d) Experimental fluorescence images of the intensity evolution and near-field images of the output facet after cleaving the device at $z_{f}=94mm$ from light injected into the $1^{st}$, $2^{nd}$, $18^{th}$ and $19^{th}$ waveguide element, and (e-h) the corresponding theoretical dynamics.}
\label{f3}
\end{center}
\end{figure}

A notable difference between an actual spin chain experiment and our optical implementation lies in the exchange symmetry of the excitations. In the present case we are dealing with bosonic entities whereas the excitations of a spin Hamiltonian are fermionic in nature. 
\begin{figure}[t!]
\begin{center}
\includegraphics[width=3in]{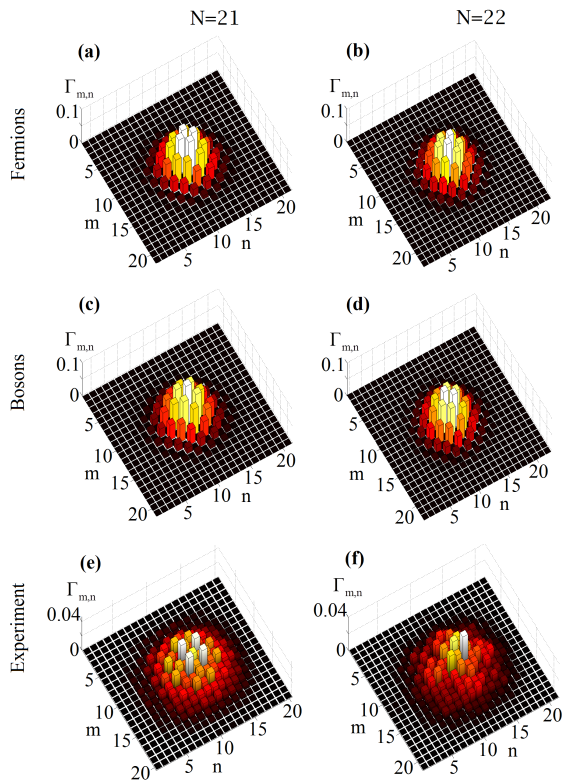}
\caption{(Color online) Correlation matrices $\Gamma_{m,n}$ corresponding to lattice systems having an odd (left) or even (right) number of elements. Theoretical comparison between (a), (b) fermionic and (c), (d) bosonic correlations when the two edge sites are excited. (e), (f) Experimentally obtained correlations using classical randomized sources emulating separable single photon states injected at the two edge sites of these arrays. All results are obtained at $z_{f}/2$.}
\label{f4}
\end{center}
\end{figure}
This fundamental difference is reflected in the arrival statistics of multi-particle experiments. To this end, we examine the correlation function $\Gamma_{n,m}=\langle a_{m}^{\dagger}a_{n}^{\dagger}a_{n}a_{m}\rangle$ \cite{Bromberg} which measures the probability for a pair of excitations arriving on site $m$  and  $n$, with each one being initialized at the extreme channels $1$ and $N$. We here focus our attention on a distance corresponding to half the state transfer length $z_{f}/2$ , i.e., when the effect is most marked since both excitations are expected to ``collide". The qualitative pattern of the correlation distribution depends on the parity of the chain (whether N is even or odd), therefore we compare the cases $N=21,$ $22$. Figures 4 (a), and 4(b) present the calculated correlations for a fermionic spin chain, whereas the corresponding results for photons are shown in figure 4(c), 4(d).  As clearly visible, the only difference in their respective correlations lies in their exchange statistics: Spin excitations can only be registered in output configurations where the difference of their positions   is odd

\begin{equation}
\label{eq444}
\Gamma_{m,n}=
\left\{
\begin{array}{ll}
\frac{1}{2^{2N-4}}{N-1\choose m-1}{N-1\choose n-1} &  n-m : odd \\
0 &  n-m : even,
\end{array}
\right.
\end{equation}

whereas for bosons it must be even

\begin{equation}
\label{eq4444}
\Gamma_{m,n}=
\left\{
\begin{array}{ll}
0 &  n-m : odd \\
\frac{1}{2^{2N-4}}{N-1\choose m-1}{N-1\choose n-1} &  n-m : even.
\end{array}
\right.
\end{equation}

Experimentally, the bosonic interference can be emulated by the interference of classical light beams with a random relative phase \cite{Keil}. In our setup, we launch two mutually coherent laser beams of equal amplitude and random relative phase into the waveguides and, and measure the classical intensity correlation or degree of second-order coherence $\Gamma_{n,m}^{c}=\Gamma_{n,m}+I_{n,1}I_{m,1}+I_{n,N}I_{m,N}$, where $I_{n,1}$ is the intensity of light output from waveguide $n$ when input to waveguide $1$  and so on. The last two terms in this expression can be experimentally measured and subtracted in order to obtain the bosonic correlation matrix $\Gamma_{n,m}$. While these two-photon correlations are independent of the phase $\Phi$, the intensities $I$ are not, and the sought after correlation can be observed only after averaging over $\Phi$ \cite{Keil}. In this experiment, we used coherent light with $\lambda=800nm$ and averaged over $60$ realizations of $\Phi$. The results of this experiment are depicted in Fig. 4(e), and 4(f), where the statistics clearly reflect the bosonic nature of the excitations used.

In conclusion we have shown that by appropriately exploiting the internal quantum dynamics of such a spin-inspired optical lattice, quantum states can be coherently transported across the functional region of an information processing device. This in turn yields significant advantages over previous strategies and provides an essential cornerstone for developing larger quantum computing devices. In this experimental work, we have explored the general concept of corruptionless quantum state transfer and we have demonstrated a high fidelity transfer through a large chain. Our results indicate that perfect state transfer protocols can provide a promising avenue towards distributed and integrated quantum computing on a chip.

A.S. thanks the German Federal Ministry of Science ad Education (Center for Innovation Competence Program, grant No. 03Z1HN31) and the German Research Foundation. R.K. is supported by the Abbe School of Photonics. A.K., L-C,K., and B.M.R.L. acknowledge support from the National Research Foundation and Ministry of Education of Singapore.

\renewcommand{\theequation}{A-\arabic{equation}}  
\setcounter{equation}{0}  
\setlength{\parindent}{0in} 
\section*{APPENDIX A: Implementation of the waveguide array and the coupling distribution}  

\begin{figure}[h!]
\begin{center}
\includegraphics[width=3.18in]{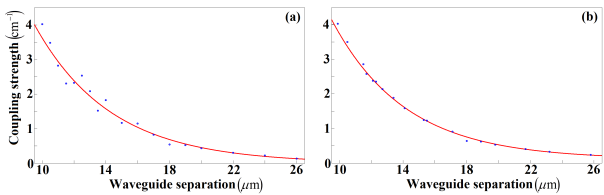}
\caption{(Color online) Measured coupling-distance dependence used for the state transfer experiment at $633nm$. The data points in (a) show the waveguide separations as programed into the fabrication stage, while the points in (b) were obtained by measuring the actual positions with a microscope.}
\label{f5}
\end{center}
\end{figure}

The waveguide lattices were inscribed using femtosecond laser techniques \cite{Szameit}. To produce the state transfer array sample, the coupling parameters must be determined. To this end, a series of optical directional couplers were fabricated using laser pulse energy of 200nJ, at a repetition rate of 100kHz, using a pulse duration of 140fs and a writing velocity of 90mm/min. Fig.\ref{f5}(a) shows the measured coupling strengths $J_{n}$ at a wavelength of $\lambda =633nm$ and for light polarization in the chip-plane vs. the inter-waveguide distance $d_{n}$ programmed into the positioning system. We fitted this dependence with the following exponential distribution $J_{n}=J_{1}\exp\left(-[d_{n}-d_{1}]/\kappa\right)$, where the required constants were found to be $d_{1}=18.1\mu m$ and $\kappa=4.81\mu m$ at $J_{1}=0.67cm^{-1}$. The above value for $J_{1}$ was chosen to yield an ideal transfer in a waveguide lattice with $N=19$ elements and of length $L=z_{f}=10cm$. The distance distribution $d_{n}=d_{1}-\kappa\ln\left(\sqrt{n\left(N-n\right)/\left(N-1\right)}\right)$ was then imposed on the transfer lattice fabricated with the same parameters in order to obtain the required coupling distribution for the \emph{spin} photonic lattice.    

\renewcommand{\theequation}{B-\arabic{equation}}  
\setcounter{equation}{0}  
\section*{APPENDIX B: Error analysis}  
The experimental setup which is used to inscribe the waveguides in silica (which in the actual transfer lattice will be separated by distances within the range of $14\mu m - 18 \mu m$) has a positioning accuracy of about $0.5\mu m$. This affects our results in two stages: The inter-waveguide spacing of the directional couplers used to obtain the coupling-distance dependence deviates from the intended values resulting in slightly biased distance parameters for the fabrication process. 
\begin{figure}[b!]
\begin{center}
\includegraphics[width=3.18in]{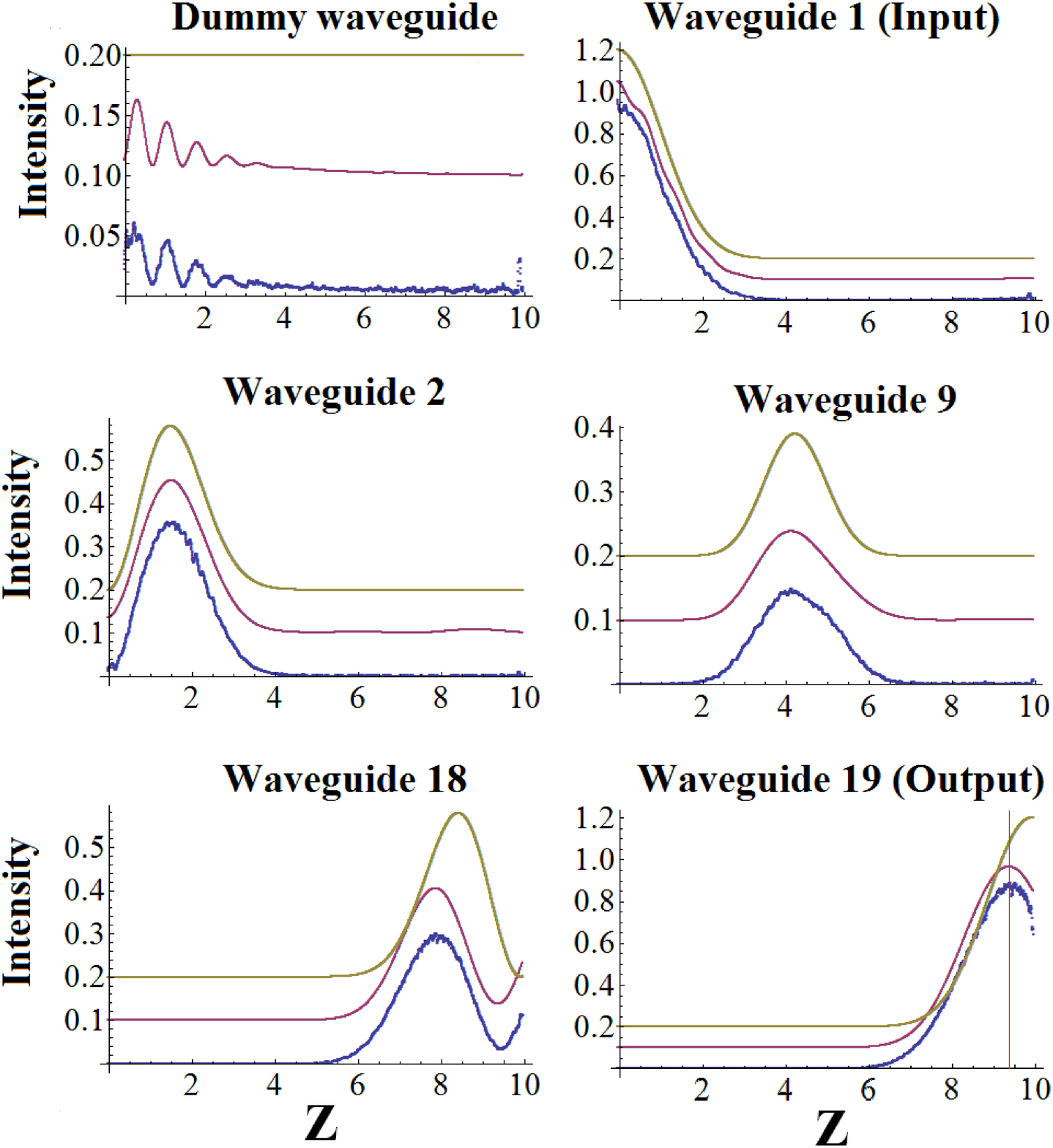}
\caption{(Color online) Comparison of experimental data (lower curve) with theoretical predictions for both an ideal system (upper curve) and the one produced (middle curve, utilizing post-production identification of system parameters) in selected waveguides. In each case, intensity is plotted as a function of the position along the sample $(Z)$, which corresponds to time in a spin chain. Light was injected in waveguide $1$. The intensity is normalized, with unity indicating that all the light is contained in the corresponding waveguide. The vertical line depicted on the second column, lowest
row symbolizes the optimum transfer distance $z_{f}=94mm$. Note that in every case the middle curve and the upper curve have an offset of 0.1 and 0.2, respectively.}
\label{f6}
\end{center}
\end{figure}
Secondly, the waveguide positions in the state transfer lattice deviate again from the calculated values, thus affecting the coupling. Post-manufacture, the waveguides can be examined under a microscope, and the true separations can be determined more accurately. For the directional couplers, this yields a clean exponential fit for the coupling vs distance dependence (Fig.\ref{f5}(b)). With these updated values one can determine the fitting values  $\kappa=4.63\mu m$ and $d_{1}=18\mu m$ at $J_{1}=0.67cm^{-1}$. Given that we implemented our system based on these values, this leads to a reduction in fidelity of approximately $5.7\%$. The accuracy in positioning the waveguides in the state transfer lattice itself is shown in Table \ref{tb:B-1}, which compares the intended separations with those produced. This effect is associated with an additional error of 4.6\%. With this understanding of the occurring positioning imperfections, amounting to a total fidelity reduction of about 10\%, we expect that in principle these effects could be compensated in future attempts. An examination of the waveguide separation indicates that, within the measurement precision of $0.2\mu m$, the spacing between the waveguides did not change along the  $Z$-length of the sample (which would translate into a time-varying coupling strength).\\ 
Due to fabrication induced stress fields in the host material, the outermost waveguides of the array were found to exhibit slightly different coupling properties. In order to minimize these effects, we inscribed one additional, 'dummy', waveguide at either end of the array, which was significantly detuned from the other waveguides. Hence, the interaction with these dummy elements was in fact negligible. From fluorescence readings, we were able to extract the degree of detuning to be $8.7cm^{-1}$, which is much larger than the coupling strengths in the lattice, but nevertheless finite. In isolation, this remaining interaction reduces the transfer fidelity by 2\%. The laser used to illuminate the waveguides had a Gaussian profile. Therefore, while most of the incident light impinged on the intended waveguide, a fraction was incident on neighboring waveguide channels, also affecting the fidelity of the system. A theoretical fit of the experimental results shows that this amounts for 4.7\% of the incident light. We emphasize that this is not a fundamental limitation of the state transfer system itself, but a practical issue related to preparing the initial state. These three effects combined account for about 17\% loss in fidelity for an excitation of the boundary and are the major sources of imperfections in our system. \\ Our choice of the $633nm$ wavelength laser enabled the use of fluorescence schemes in order to observe from the top the light intensity along the length of the sample, instead of just detecting it at the output. While this benefits our comparison to the theoretical results (including the error analysis), it also implies that a substantial amount of photon loss is present throughout the sample. 
\begin{center}
\begin{table*}[h!]
{\small
\hfill{}
\begin{tabular}{|c|c|c|c|c|c|c|c|c|c|c|c|c|c|c|c|c|c|c|c|c|c|c|}
\hline
Left-most waveguide&Dummy & 1 & 2 & 3 & 4 & 5 & 6 & 7 & 8 & 9\\
\hline
$d_{I}$ & 17 & 18.12 & 16.59 & 15.76 & 15.22 & 14.85 & 14.59 & 14.42 & 14.3 & 14.25\\
\hline
$d_{F}$ & 16.6 & 18.3 & 18.8 & 15.8 & 14.6 & 14.3 & 15 & 14.3 & 14 & 14\\
\hline
Left-most waveguide & 10 & 11 & 12 & 13 & 14 & 15 & 16 & 17 & 18 & 19\\
\hline
$d_{I}$ & 14.25 & 14.3 & 14.42 & 14.59 & 14.85 & 15.22 & 15.76 & 16.59 & 18.12 & 17\\
\hline
$d_{F}$ & 14.9 & 13.7 & 14 & 14.9 & 15.1 & 14.5 & 15.1 & 16.7 & 18.1 & 17.3\\
\hline
\end{tabular}}
\hfill{}
\caption{Comparison of intended waveguide separations $(d_{I})$ on the state transfer system with those actually fabricated $(d_{F})$. All distances are in $\mu m$.}
\label{tb:B-1}
\end{table*}
\end{center}
In subsequent data processing, we have therefore renormalized the system so that the light intensity per unit distance in the $Z$ direction is constant. The post-selection upon arrival does not detract from the realization of the spin chain which is meant to be lossless. However, the fluorescence information exhibits saturation effects (particularly at large intensities) and is relatively susceptible to background noise, making the data obtained from the near-field images of the intensity pattern at the end of the sample the most reliable for calculating transfer fidelities. From the output of the $100mm$ long sample, we have initially obtained a transfer fidelity of 76\% from port   to port   (74\% in the reverse direction).  \\ Timing errors can also potentially have a large impact - the arrival intensity in the outermost waveguides (in the ideal case) can be expressed as 
\begin{equation}
\label{B1}
F_{1N}=F_{N1}=\left[\sin\left(\frac{\pi Z}{2z_{f}}\right)\right]^{2(N-1)},
\end{equation}
which is tightly focused at $z_{f}$ for large $N$. Having originally made the sample slightly too long, we were able to cut back along $Z$  in order to find the optimal point of transfer, at $94mm$ (see Fig.\ref{f6}). 
Cleaving the sample, and measuring the output intensities there, yielded an improved transfer fidelity of 82\% for the transfer occurring from site $1$  to $N=19$  whereas in the opposite direction the fidelity reached 84\%. This illustrates one of the many benefits of pre-manufacturing a state transfer chain rather than dynamically generating the same effect. In other words one can perform these tests and determine the optimum length in view of the other experimental imperfections that have arisen in the system. In Fig.\ref{f6}, we compare the (renormalized) experimental data with the simulated evolution based on subsequent measurements that determined more accurately the positioning of the waveguides and their interaction strength. Clearly, this good agreement bodes well for future experiments in which these errors can be better controlled. Indeed, our theoretical model suggests that even in the current system, with perfect initial state preparation, at the optimum length, it will be possible to achieve a fidelity in excess of 93\%.
\bibliography{Jx}

\begin{thebibliography}{32}
\bibitem[{fid()}]{fidelity}
 (????), \bibinfo{note}{here the fidelity is defined as the optical intensity
  $F\equiv |\bra{p}\exp\left(iZJ_{x}\right)\ket{q}|^{2}=|T_{p,q}|^{2}$}.
\end{thebibliography}
\end{document}